\documentclass[conference]{IEEEtran}
\IEEEoverridecommandlockouts
\usepackage{cite}
\usepackage{amsmath,amssymb,amsfonts}
\usepackage{algorithmic}
\usepackage{graphicx}
\usepackage{textcomp}
\usepackage{xcolor}
\usepackage{float}
\usepackage{booktabs}
\usepackage{array}
\usepackage{placeins}
\usepackage{hyperref}
\usepackage[T1]{fontenc}
\usepackage[utf8]{inputenc}
\usepackage[serbian,english]{babel}

\usepackage{pgfplots}
\usepackage{pgfplotstable}

\usepackage{algorithm}
\usepackage{algorithmic}

\usepackage{amsthm}

\theoremstyle{definition}
\newtheorem{definition}{Definition}[section]  

\usepackage{amsthm}

\def\BibTeX{{\rm B\kern-.05em{\sc i\kern-.025em b}\kern-.08em
    T\kern-.1667em\lower.7ex\hbox{E}\kern-.125emX}}
\begin{document}

\title{GGBond: Growing Graph-Based AI-Agent Society for Socially-Aware Recommender Simulation\\
{\footnotesize \textsuperscript{}}
\thanks{}
}

\author{
\IEEEauthorblockN{
Hailin Zhong\textsuperscript{1}, 
Hanlin Wang\textsuperscript{1}, 
Yujun Ye\textsuperscript{1}, 
Meiyi Zhang\textsuperscript{1}, 
Shengxin Zhu\textsuperscript{2,1}
}
\IEEEauthorblockA{
\textsuperscript{1}\textit{Faculty of Science and Technology, Beijing Normal-Hong Kong Baptist University}, Zhuhai, China\\
\textsuperscript{2}\textit{Research Centers for Mathematics, Advanced Institute of Natural Sciences, Beijing Normal University}, Zhuhai, China\\
\{r130026215, r130026135, r130026180, r130026169\}@mail.uic.edu.cn, shengxinzhu@uic.edu.cn
}
}

\maketitle
\vspace{-6em}  

\begin{abstract}
Current personalized recommender systems predominantly rely on static offline data for algorithm design and evaluation, significantly limiting their ability to capture long-term user preference evolution and social influence dynamics in real-world scenarios. To address this fundamental challenge, we propose a high-fidelity social simulation platform integrating human-like cognitive agents and dynamic social interactions to realistically simulate user behavior evolution under recommendation interventions. Specifically, the system comprises a population of Sim-User Agents, each equipped with a five-layer cognitive architecture that encapsulates key psychological mechanisms, including episodic memory, affective state transitions, adaptive preference learning, and dynamic trust-risk assessments. In particular, we innovatively introduce the Intimacy–Curiosity–Reciprocity–Risk (ICR²) motivational engine grounded in psychological and sociological theories, enabling more realistic user decision-making processes. Furthermore, we construct a multilayer heterogeneous social graph (GGBond Graph) supporting dynamic relational evolution, effectively modeling users' evolving social ties and trust dynamics based on interest similarity, personality alignment, and structural homophily. During system operation, agents autonomously respond to recommendations generated by typical recommender algorithms (e.g., Matrix Factorization, MultVAE, LightGCN), deciding whether to consume, rate, and share content while dynamically updating their internal states and social connections, thereby forming a stable, multi-round feedback loop. This innovative design transcends the limitations of traditional static datasets, providing a controlled, observable environment for evaluating long-term recommender effects.
\end{abstract}


\section{Introduction}
In recent years, personalized recommender systems have become widely deployed across digital platforms to alleviate information overload and deliver precise content matching. While such systems have significantly improved user experience and platform performance, mainstream research and evaluation practices remain heavily dependent on static historical logs. These static datasets provide only a snapshot of user behaviors, failing to reflect how user preferences evolve over long-term interactions or how social influences shape recommendation acceptance. Prior work has shown that ignoring such dynamics can lead to short-sighted optimization, filter bubbles, and even polarization effects in recommender outputs. Thus, enabling recommendation algorithms to understand, adapt to, and be evaluated under long-term behavioral dynamics and social interactivity remains a fundamental yet unresolved challenge in the field. 

To address these limitations, several simulation-based evaluation frameworks have been proposed (e.g., RecSim), aiming to test recommendation algorithms beyond static data. However, most of these simulators model users as reward-driven policy executors, lacking the ability to emulate realistic human decision-making processes that involve memory, emotion, evolving preferences, and social cognition. Meanwhile, the majority of social recommendation models still rely on static adjacency graphs, which overlook key sociological features such as tie-strength drift, trust evolution, multi-context social layers, and indirect influence propagation \cite{tang2013social}. These simplifications critically hinder our ability to study recommender systems in real-world interactive and socially complex environments.

\begin{figure*}
    \centering
    \includegraphics[width=1\linewidth]{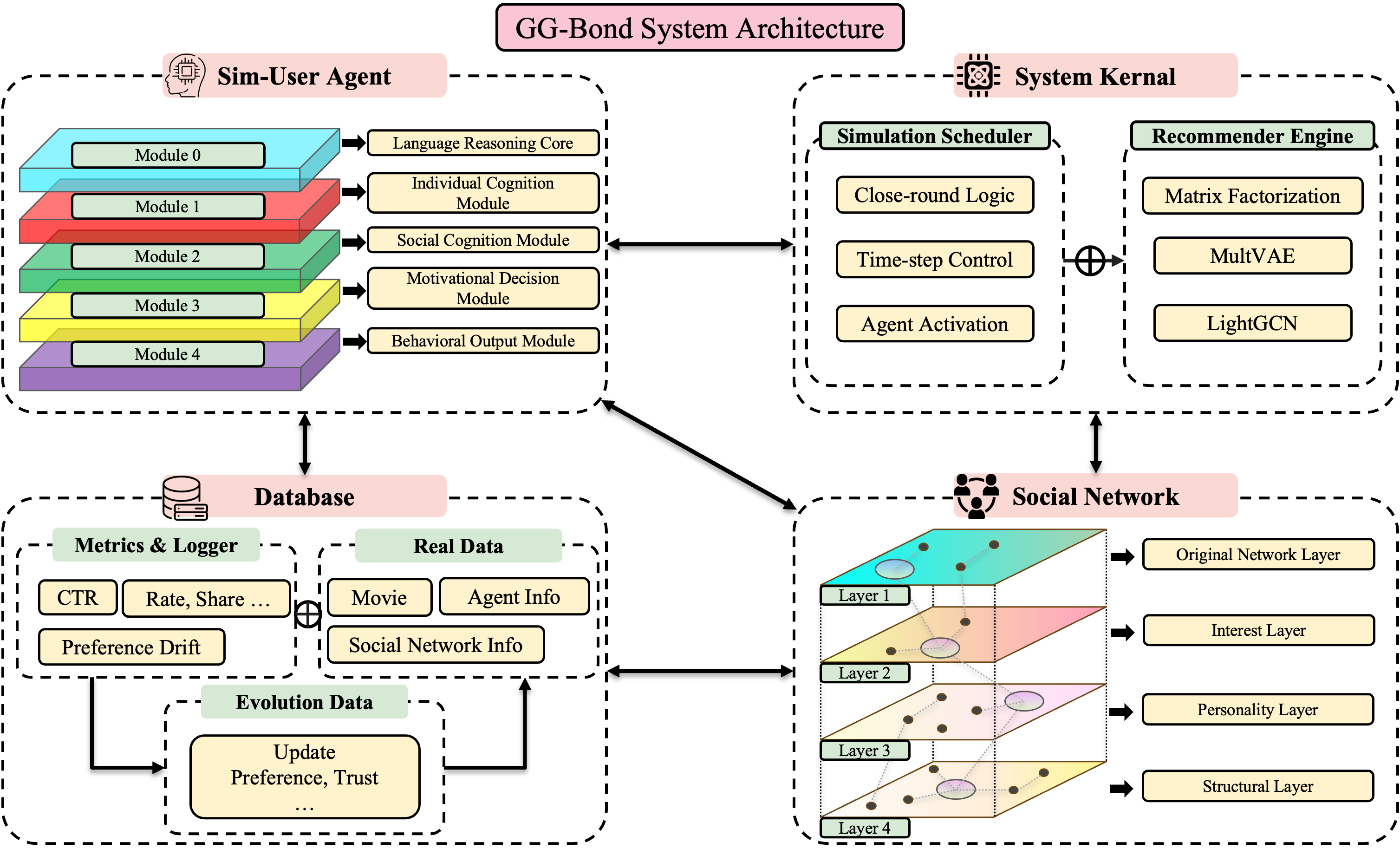}
    \caption{GGBond System Architecture: Rrecommender Engine, Database, Social network, Agent}
    \label{fig:system}
\end{figure*}

To bridge these gaps, we argue for the construction of a high-fidelity simulation environment that jointly models cognitive user behavior, heterogeneous social structure, and causal feedback loops. In such an environment, each user should demonstrate rich internal decision-making behavior—affected by memory, affective state, and evolving interests—while simultaneously engaging in social interactions guided by intimacy, trust, and reciprocity. These interactions, in turn, should influence content diffusion and preference updates over time. A simulation platform of this nature would not only provide an expressive substitute for limited real-world datasets but also enable rigorous testing of recommendation algorithms under complex dynamics and long-term feedback. 

In this work, we propose a novel high-fidelity simulation platform \textit{\textbf{Growth Graph-based Recommendation System via AI-Agent Social Bonds (GGBond)}} for recommender systems that models both user cognition and social interaction. The platform consists of two core components: (1) a population of simulated user agents with psychologically grounded cognitive architectures, and (2) a dynamically evolving multi-layer social graph (GGBond Graph) that supports heterogeneous semantics and multi-dimensional social relations. Each simulated agent is endowed with a five-layer architecture that includes memory, emotion, preferences, trust assessment, and natural language generation, enabling fine-grained modeling of internal decision processes. The decision of whether to consume and share content is governed by a psychologically-inspired motivational engine (ICR²), which integrates intimacy ($I$), novelty ($N$), reciprocity potential ($R$), and risk perception ($K$) into a unified decision score. On the structural side, we enrich an anonymized real-world topology (e.g., Stanford Facebook graph) with inferred personality vectors using a structure-driven inference model, and construct layered graphs capturing interest similarity, personality alignment, and structural homophily. The system is executed under a discrete-time simulation engine. In each time step, a recommendation model (e.g., Matrix Factorization, MultVAE, or LightGCN) generates candidate content based on the current state of users and the social graph. Simulated agents autonomously decide whether to watch, rate, and share the content. These behaviors trigger updates to the agent’s internal state and modify the social graph accordingly. All behavioral signals and structural dynamics are recorded, enabling multifaceted evaluation of recommender algorithms on metrics such as preference drift, diversity, and social influence. Additionally, these records serve as empirical validation of the simulation’s behavioral realism.

Compared to existing approaches, our proposed system makes the following structural contributions: 
\begin{itemize}
\item We introduce a simulation paradigm that integrates human-like cognition and social evolution, enabling long-term recommender evaluation beyond static logs; 
\item We design a psychologically grounded agent architecture incorporating memory, affect, preference, and a novel ICR² motivational engine to simulate realistic user decisions; 
\item We construct a heterogeneous, multi-context, and dynamically updating social graph that reflects evolving intimacy and trust relationships; and 
\item We provide an extensible evaluation platform that supports the integration of representative recommendation algorithms and enables systematic analysis of their performance and social impact in realistic interactive settings. 
\end{itemize}

\section{GGBond System Architecture}
\label{sec:system}
To simulate user behavior in a socially contextualized recommendation environment with high fidelity, we design a virtual society framework composed of a population of \textit{Sim-User Agents} and a multi-layer heterogeneous social graph called the \textit{GGBond Social Network}. The system architecture is clearly presented in Figure~\ref{fig:system}. This system is built not only to reproduce realistic cognitive and behavioral patterns of individuals under social influence, but also to serve as a controlled, repeatable testbed for evaluating the performance of mainstream recommendation algorithms in complex user scenarios.

During initialization, we construct an anonymized social graph based on the Stanford Facebook dataset and enrich each node with Big-Five personality traits through a structure-driven inference framework. The resulting hybrid profile integrates structural features and psychological traits. On top of this, we define a multi-layer social network consisting of interest-based, personality-based, and structural homophily layers, all of which are fused into a unified layer for downstream reasoning. This network supports layered semantic modeling and serves as the basis for trust estimation, social propagation, and motivational reasoning.

Each agent operates through a five-Module cognitive architecture. The language reasoning core (Module 0) generates natural language outputs for commentary and interaction; the individual cognition Module (Module 1) models memory, affective state, and evolving content preferences; the social cognition Module (Module 2) maintains multi-dimensional intimacy scores and trust estimations toward other agents; the motivational engine (Module 3) computes the internal drive $C$ based on intimacy ($I$), novelty ($N$), reciprocity potential ($R$), and risk perception ($K$); and the behavior output Module (Module 4) executes watch, rate, and share actions, writing feedback back into upper Modules, forming a complete perception–decision–action loop.

The system runs on a discrete time-step simulation engine. Each time step corresponds to an actionable social or system-triggered recommendation opportunity. A scheduler activates agents sequentially, allowing them to perceive incoming content, evaluate it, and autonomously decide whether to consume, rate, and propagate it. This asynchronous update scheme better reflects real-world behavioral heterogeneity than synchronous models and avoids unrealistic global coordination assumptions.

Recommendation algorithms are embedded within this evolving simulation. At each time step, algorithms take as input the current state of the agents and social graph, generating personalized candidate recommendations. Agents, in turn, respond to these recommendations according to their current internal states. This interaction forms a closed evaluation loop, allowing us to monitor both traditional metrics (e.g., CTR, acceptance rate, diversity) and long-term behavioral consequences (e.g., preference drift, social influence spread). The system supports both periodic algorithmic interventions and context-sensitive triggers, such as increased activity periods or mismatched preference stages.

Throughout the simulation, we collect a wide range of structural and behavioral metrics, including graph dynamics (e.g., density, community shifts), internal agent state distributions (e.g., affect variation, preference entropy), and recommendation interaction traces. A built-in write-back mechanism ensures that every decision made by the agent has forward-propagating consequences, enabling systematic analysis of long-term recommender interventions.

The system currently supports three representative recommendation algorithms: Matrix Factorization, MultVAE, and LightGCN. These algorithms operate directly on the evolving user profiles and social graph, allowing for unified evaluation across personalization quality, social robustness, and adaptation under long-term user drift. This dual evaluation paradigm not only benchmarks algorithmic performance in realistic environments but also serves as indirect validation of our simulation framework’s behavioral and structural credibility.

Through the coordinated design of internal cognitive agents and an external multi-layer social network, the system offers a scalable, interpretable, and behaviorally grounded simulation platform for research in personalized recommendation, social trust modeling, and emergent group behavior dynamics.

\section{Social Network Architecture}

This study presents an advanced social network architecture, meticulously crafted to synthesize anonymized structural attributes from the Stanford Facebook social graph \cite{Stanford} with nuanced personality profiles inferred from external behavioral datasets, such as MovieLens \cite{harper2015movielens}, Steam \cite{kang2018self}, and AmazonBooks \cite{mcauley2015image}, which offer rich textual review corpora for psychological analysis \cite{schwartz2013personality}. The fundamental challenge underpinning this integration arises from the inherent anonymization of the Stanford dataset—core demographic attributes including occupation, educational background, and age are withheld, effectively precluding direct personality alignment with external datasets.

\begin{figure}
    \centering
    \includegraphics[width=1\linewidth]{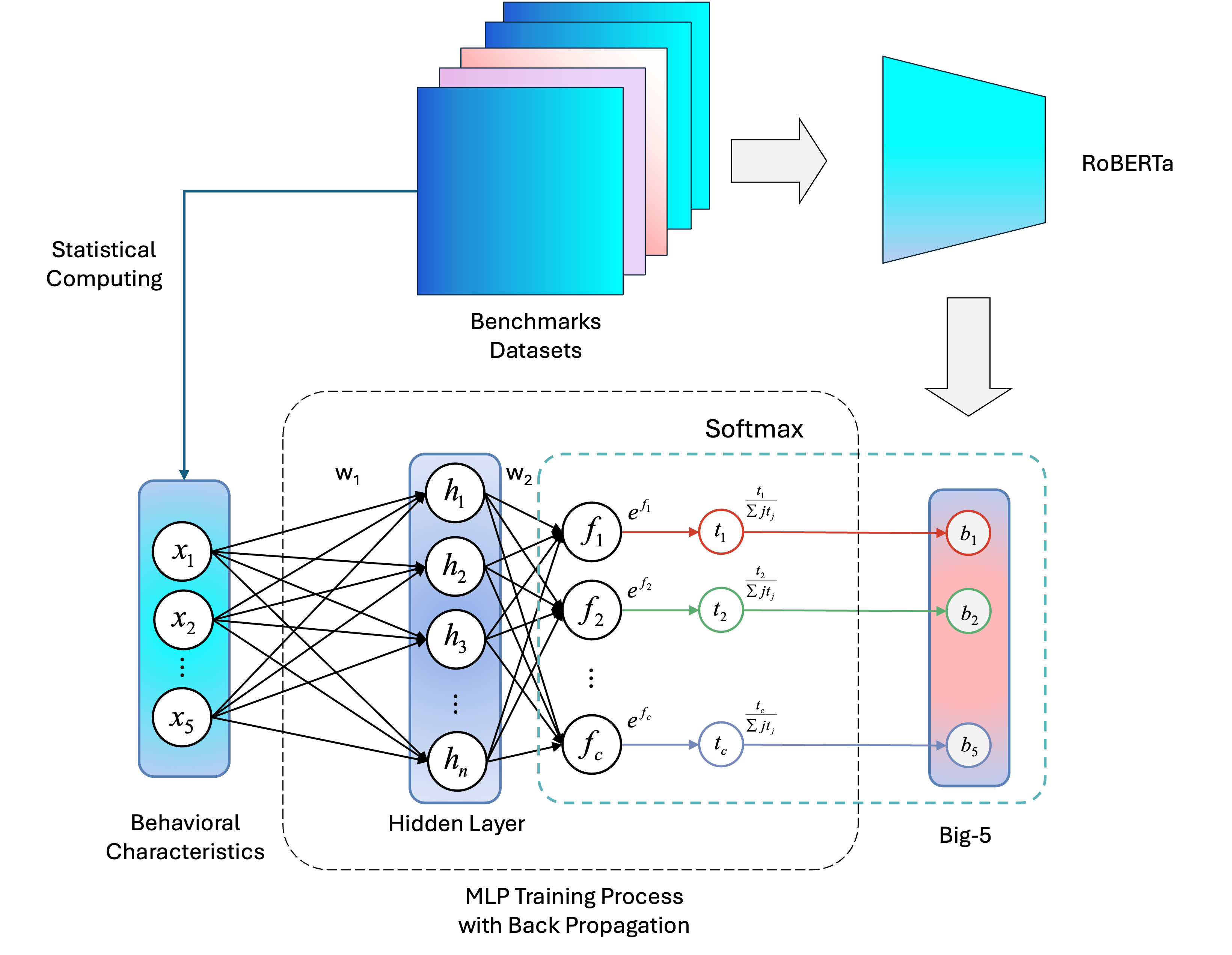}
    \caption{Big-5 Model Training Process}
    \label{fig:training}
\end{figure}

\begin{figure}
    \centering
    \includegraphics[width=\linewidth]{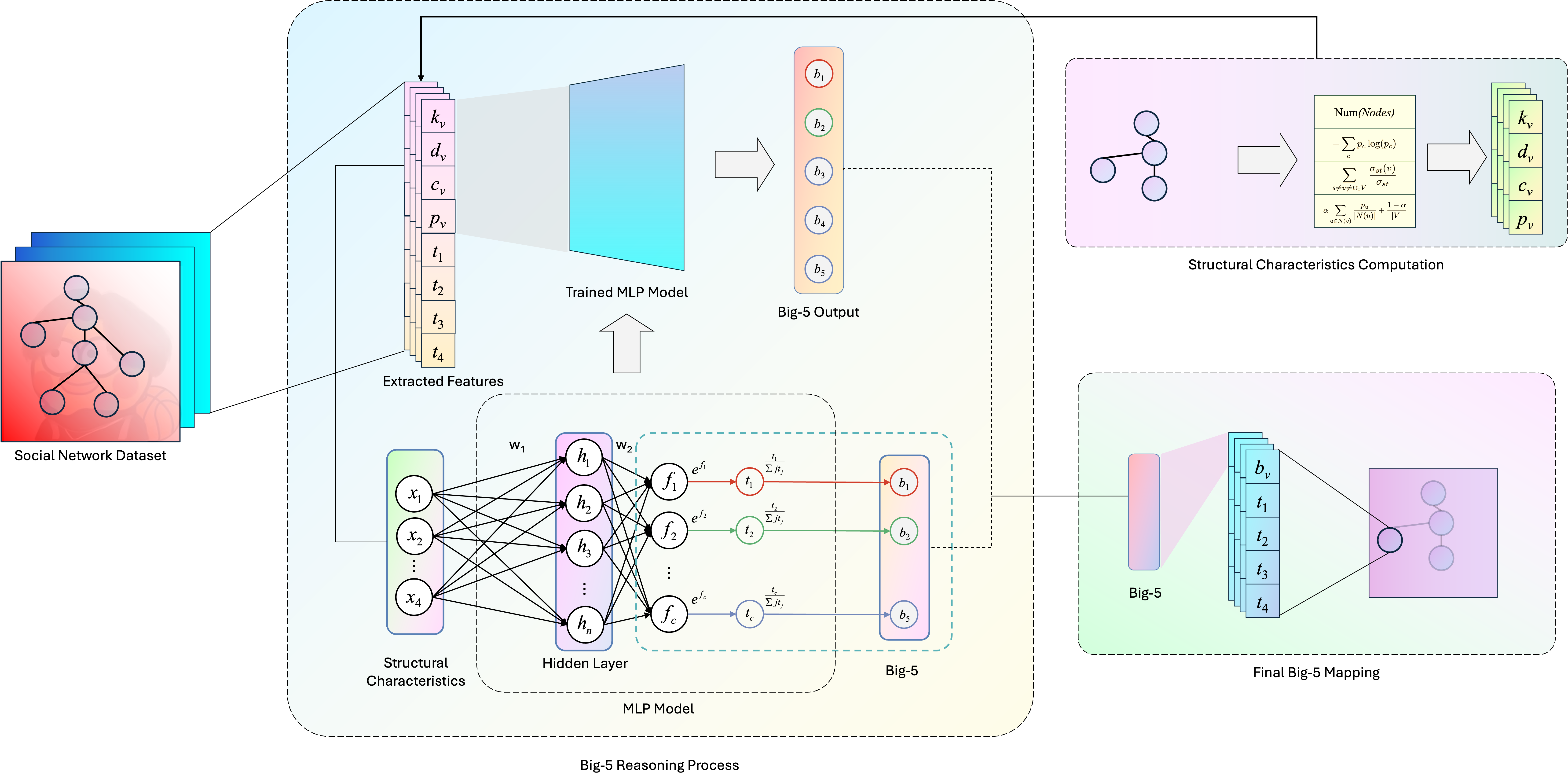}
    \caption{Big-5 Alignment Framework}
    \label{fig:process}
\end{figure}

To navigate this complexity, we propose a robust two-stage, structure-driven inferential framework that capitalizes solely on the intrinsic topology of the social graph to predict latent Big-Five personality traits for each node. In the first stage (illustrated in Figure \ref{fig:training}), we leverage textual reviews from AmazonBooks, Steam, and MovieLens datasets to extract Big-Five personality traits using a pre-trained RoBERTa model. Simultaneously, we derive behavioral statistics from user interactions in MovieLens, including activity level, diversity, rating deviation, and novelty preference. In the second stage (shown in Figure \ref{fig:process}), the trained MLP model is applied to the Stanford network. Here, structurally analogous features are extracted solely from network topology—node degree for social activity, entropy-based neighbor diversity for social diversity, betweenness centrality \cite{freeman1977set} for deviation from mainstream, and PageRank scores \cite{page1999pagerank} for novelty preference. These features serve as proxies for traditional psychological indicators within the social graph context \cite{kosinski2013private}. The learned mapping enables the prediction of Big-Five traits for anonymized nodes.

\subsection{Personality Prediction Model}

To predict the Big-Five personality traits (Openness, Conscientiousness, Extraversion, Agreeableness, Neuroticism), we train a multi-task regression model using behavioral data extracted from the MovieLens dataset. Each user is represented by four interpretable features grounded in their historical interactions:

\begin{definition}[Activity Level]
Activity Level $T_u^{\text{act}}$ denotes the total number of items rated by user $u$, capturing their engagement intensity:
\begin{equation}
T_u^{\text{act}} = \sum_{i \in I_u} 1,
\end{equation}
where $I_u$ is the set of items rated by user $u$.
\end{definition}

\begin{definition}[Diversity]
Diversity $T_u^{\text{div}}$ measures the entropy of genre distribution among all rated items, reflecting the breadth of a user’s interests:
\begin{equation}
T_u^{\text{div}} = - \sum_{g \in G} p_{u,g} \log p_{u,g}, \quad p_{u,g} = \frac{|\{ i \in I_u \mid \text{genre}_i = g \}|}{|I_u|},
\end{equation}

where $G$ is the set of all genres.
\end{definition}

\begin{definition}[Conformity Deviation]
Conformity Deviation $T_u^{\text{conf}}$ quantifies the deviation of user $u$’s ratings from the global consensus:
\begin{equation}
    T_u^{\text{conf}} = \frac{1}{|I_u|} \sum_{i \in I_u} (r_{ui} - \bar{R}_i)^2,
\end{equation}

where $r_{ui}$ is the rating given by user $u$ to item $i$, and $\bar{R}_i$ is the average rating for item $i$ across all users.
\end{definition}

\begin{definition}[Novelty Seeking]
Novelty Seeking $T_u^{\text{nov}}$ captures the user’s tendency to interact with less popular (i.e., rare) items:
\begin{equation}
    T_u^{\text{nov}} = \frac{1}{|I_u|} \sum_{i \in I_u} \frac{1}{\text{pop}(i)}, \quad \text{pop}(i) = |\{ u' \mid i \in I_{u'} \}|,
\end{equation}

where $\text{pop}(i)$ is the popularity of item $i$, defined as the number of users who rated it.
\end{definition}

This trained model is subsequently applied to structural features derived from the Stanford social graph, enabling cross-domain personality prediction for anonymized nodes.

\subsection{Structure-based Feature Engineering}

\begin{definition}[Social Activity]
The Social Activity ($k_v$) is defined as the node degree, representing the number of direct connections a node has in the social graph, analogous to user activity level. 
\end{definition}

\begin{definition}[Social Diversity]
The Social Diversity ($d_v$) is computed as the entropy of anonymized occupation or education labels among a node’s immediate neighbors, representing the variety in a node’s social contacts:
\begin{equation}
    d_v = -\sum_{c} p_{c}\log(p_{c}),
\end{equation}
where $p_c$ denotes the proportion of neighbors belonging to category $c$.
\end{definition}

\begin{definition}[Deviation from Mainstream]
The Deviation from Mainstream ($c_v$) is captured via betweenness centrality, measuring the extent to which a node acts as a bridge between distinct social communities, thus indicating non-conformity:
\begin{equation}
    c_v = \sum_{s \neq v \neq t \in V} \frac{\sigma_{st}(v)}{\sigma_{st}},
\end{equation}
where $\sigma_{st}$ is the total number of shortest paths between nodes $s$ and $t$, and $\sigma_{st}(v)$ is the number of these paths passing through node $v$.
\end{definition}

\begin{definition}[Novelty Preference]
The Novelty Preference ($p_v$) is measured by the node’s PageRank score, reflecting a node's position relative to network prominence, with lower scores indicating a preference towards peripheral or less mainstream connections:
\begin{equation}
    p_v = \alpha \sum_{u \in N(v)} \frac{p_u}{|N(u)|} + \frac{1-\alpha}{|V|},
\end{equation}
where $N(v)$ is the set of neighbors of node $v$, and $\alpha$ is the damping factor (typically 0.85).
\end{definition}

Finally, the structural feature vector $\mathbf{x}_v$ for each node $v$ is defined as:
\begin{equation}
    \mathbf{x}_v = [k_v, d_v, c_v, p_v].
\end{equation}

\subsection{Personality Reasoning Process}

To predict the Big-Five personality traits (Openness, Conscientiousness, Extraversion, Agreeableness, Neuroticism) for Stanford nodes, we trained a multi-task regression model using data derived from Real banchmark datasets' user behaviors and personality traits extracted from textual reviews. Specifically, we utilized a pre-trained RoBERTa-based classifier on MovieLens textual data to label user personalities and trained a neural network $f_\theta$ to map structural behavior features to personality vectors:
\begin{equation}
    \hat{\mathbf{b}}_v = f_\theta(\mathbf{x}_v),
\end{equation}
where $\hat{\mathbf{b}}_v$ denotes the predicted Big-Five personality vector for node $v$.

The model was trained on real datasets features and evaluated based on the RMSE and Pearson correlation metrics on a hold-out set, achieving robust predictive performance (RMSE $<$ 0.1, Pearson $r > 0.6$). The trained model was subsequently applied to predict the personality traits of Stanford network nodes, producing personality-enriched node representations without reliance on explicit personal data.

\subsection{Big-Five Mapping}

In the Stanford social network dataset, each node $v$ is originally described by a set of anonymized social attributes and topological features, formally represented as:
\begin{equation}
    \mathbf{t}_u = [t_1, t_2, t_3, t_4],
\end{equation}
where: $t_1$ denotes the number of connected edges (degree). $t_2$ represents the anonymized occupation identifier. $t_3$ represents the anonymized age group identifier. $t_4$ represents the anonymized education identifier.

Following the Big-Five personality prediction process, we enrich this attribute vector by appending the predicted Big-Five personality traits, denoted as:
\begin{equation}
    \mathbf{b}_v = [b_1, b_2, b_3, b_4, b_5],
\end{equation}
where each $b_i$ corresponds to one dimension of the Big-Five personality (Openness, Conscientiousness, Extraversion, Agreeableness, Neuroticism), scaled within the range $[0,1]$.

The complete profile vector for node $w$ thus becomes:
\begin{equation}
    \mathbf{p}_w = [\mathbf{t}_u, \mathbf{b}_v].
\end{equation}

This unified representation integrates structural social attributes and latent psychological traits, providing a comprehensive embedding for each node. Such enriched profiles facilitate downstream tasks, including personalized recommendation and social influence modeling, by co-optimizing structural and psychological dimensions within the network.

\subsection{GGBond Social Network}

Building upon the personality-enriched node representation $\mathbf{p}_u = [\mathbf{t}_u, \mathbf{b}_u]$, where $\mathbf{t}_u \in \mathbb{R}^4$ denotes anonymized structural attributes (e.g., degree, occupation, age, education), and $\mathbf{b}_u \in \mathbb{R}^5$ represents predicted Big-Five personality traits, the GGBond framework constructs a multilayer social graph $\mathcal{G}$ to simulate user interactions, trust propagation, and preference shifts.

\begin{figure}[ht]
    \centering
    \includegraphics[width=1\linewidth]{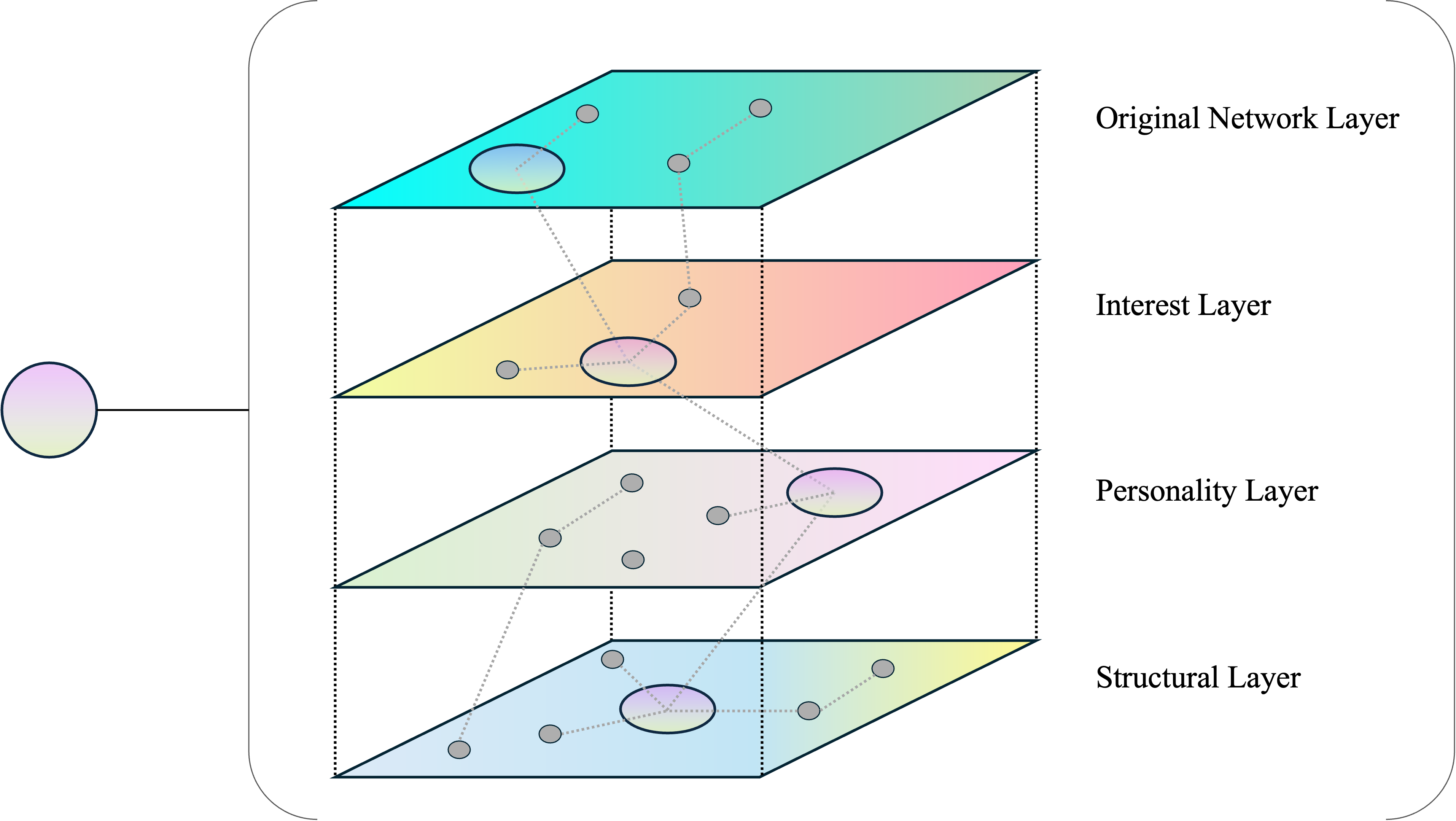}
    \caption{GGBond Multi-layer Social Network Framework. Each colored plane represents a type of social edge (Interest, Personality, Structural, or Unified), with dotted connections modeling influence across layers. Agents (nodes) are embedded in all layers simultaneously. The aggregation and propagation mechanism across these layers enables personality drift and preference adaptation.}
    \label{fig:ggbond}
\end{figure}

As visualized in Figure~\ref{fig:ggbond}, the architecture is composed of several semantically distinct graph layers, each encoding a different dimension of social similarity or homophily.

\paragraph{Interest Graph Layer.}
In the top (cyan) layer, edges are created based on user preference overlap. An undirected edge is formed between agents $u$ and $v$ if they share liked movie types, weighted by Jaccard similarity:
\begin{equation}
   w_{uv}^{\text{int}} = \frac{|\mathcal{M}_u \cap \mathcal{M}_v|}{|\mathcal{M}_u \cup \mathcal{M}_v|}, 
\end{equation}

where $\mathcal{M}_u$ is the set of positively rated movies types by $u$. And we select Top-5 similarity of each agents.

\paragraph{Personality Graph Layer.}
In the second (orange-pink) layer, agents are connected if they have similar psychological traits. The edge weight is defined using cosine similarity:
\begin{equation}
    w_{uv}^{\text{pers}} = \cos(\mathbf{b}_u, \mathbf{b}_v) = \frac{\mathbf{b}_u \cdot \mathbf{b}_v}{\|\mathbf{b}_u\| \cdot \|\mathbf{b}_v\|}.
\end{equation}

\paragraph{Structural Graph Layer.}
The third (purple-green) layer models homophily in terms of occupation and age category. Edge weights are binary, based on shared categories:
\begin{equation}
    w_{uv}^{\text{struct}} = \mathbf{1}_{\text{same}(t_{u,2}, t_{v,2})} + \mathbf{1}_{\text{same}(t_{u,3}, t_{v,3})},
\end{equation}

\paragraph{Unified Layer and Aggregation.}
The whole layer aggregates all previous edges using a weighted linear combination:
\begin{equation}
    w_{uv} = \alpha \cdot w_{uv}^{\text{int}} + \beta \cdot w_{uv}^{\text{pers}} + \gamma \cdot w_{uv}^{\text{struct}}, \quad \text{with } \alpha + \beta + \gamma = 1.
\end{equation}

This unified, weighted graph is used during social reasoning and multi-round preference simulation.

\paragraph{Multi-Round Interaction Protocol.}
During each round, agent $u$ receives candidate movie $m$ from its neighbors $N(u)$ and decides whether to accept based on semantic alignment and social propagation provided in Agent Architecture

Accepted movies are added to agent profile, and agent embeddings are updated as described in Layer~1. In addition, personality drift is also triggered by successful social influence, the formula is given in Agent Architecture section.

\paragraph{Final Output.}
After $T$ rounds, the updated agent profiles $\{\mathbf{p}_u^{(T)}\}$ and graph $\mathcal{G}^{(T)}$ are passed to downstream recommenders (e.g., MF, MultVAE, LightGCN) to generate final predictions. This layered social framework enables rich modeling of community structure, latent personality alignment, and iterative preference shaping over the social graph.

\section{Agent Architecture}
\label{sec:agent}
\begin{figure}
    \centering
    \includegraphics[width=\linewidth]{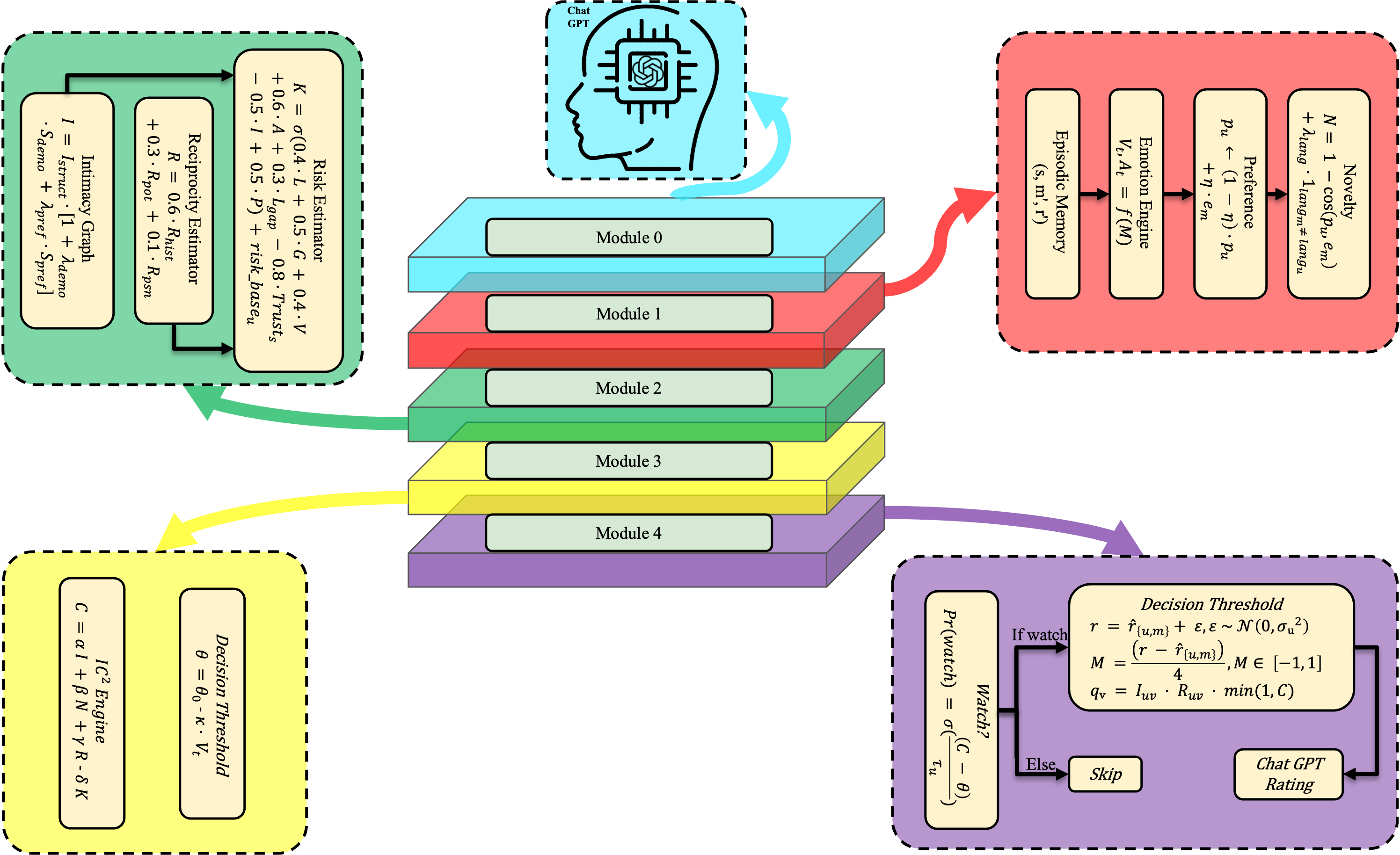}
    \caption{Agent Architecture:
    Module 0 (GPT4 API), Module 1 (Individual cognition Module), Module 2 (Social cognition Module), Module 3 (Decision Module), Module 4 (Behavior Module)}
    \label{fig:agent}
\end{figure}
The agent architecture is clearly presented in Figure~\ref{fig:agent}, each simulated agent is designed as a fully autonomous entity equipped with the capabilities of perception, cognition, decision-making, and behavioral feedback. Its actions are entirely driven by internal mechanisms, without reliance on external calls. The agent operates through a closed cognitive–motivational–behavioral loop: upon receiving an external stimulus—such as a system-generated recommendation or a peer-shared item—the agent invokes a series of internal cognition modules to encode and evaluate the environment, resulting in a set of psychologically and socially meaningful latent variables\cite{sun2006clarion}. These variables are then passed to the motivational engine, which computes internal drive and determines whether and how the agent should respond. The agent may choose to watch the movie, assign a rating, and share it with others, while also writing feedback back into its memory, emotional state, and social ties, thereby enabling long-term behavioral adaptation.

This autonomy is enabled by the coordinated operation of multiple functional subsystems, including episodic recall, affect regulation, preference learning, and social trust-risk inference. To systematically support these capabilities, we introduce a structured five-Module agent architecture that aligns with core components of behavior: language generation, individual cognition, social reasoning, motivational computation, and behavioral output. In the following sections of this architecture, each Module is described in detail, corresponding to a specific stage in the internal decision pipeline and together implementing the cognitive loop described above\cite{ie2019recsim}. We further summarize the operational logic of the agent as an algorithm~\ref{alg:agent_cycle}.

\begin{algorithm}[ht]
\caption{Simulated Agent Decision Cycle}
\label{alg:agent_cycle}
\small
\begin{algorithmic}[1]
\REQUIRE Received recommendation $m$ from source $s$ at time $t$
\ENSURE Agent response: \texttt{watch/skip}, rating $r$, share set $\mathcal{S}$

\STATE \textbf{Module 1: Individual Cognition}
\STATE Retrieve past event history $(s, m', r')$ from episodic memory
\STATE Compute current emotional state $(V_t, A_t)$ from past satisfy $M$
\STATE Update preference vector $\mathbf{p}_u$
\STATE Compute novelty score $N \gets \texttt{Novelty}(\mathbf{p}_u, \mathbf{e}_m, \text{lang})$

\STATE \textbf{Module 2: Social Cognition}
\STATE Retrieve intimacy score $I \gets \texttt{Intimacy}(u, s)$
\STATE Estimate reciprocity potential $R \gets \texttt{Reciprocity}(u, s)$
\STATE Evaluate risk perception $K \gets \texttt{Risk}(m, s, u)$

\STATE \textbf{Module 3: Motivation Evaluation}
\STATE Compute dynamic threshold $\theta \gets \theta_0 - \kappa \cdot V_t$
\STATE Compute motivation score $C \gets \alpha I + \beta N + \gamma R - \delta K$

\STATE \textbf{Module 4: Behavior Execution}
\STATE Compute watch probability $p \gets \sigma((C - \theta)/\tau_u)$
\IF{Sample($\texttt{Bernoulli}(p)$) == watch}
    \STATE Predict expected rating $\hat{r}_{u,m}$
    \STATE Sample final rating $r \gets \hat{r}_{u,m} + \varepsilon$
    \STATE Compute satisfaction $M \gets (r - \hat{r}_{u,m}) / 4$
    \STATE Select share targets $\mathcal{S} \gets \texttt{TopK}(q_v = I_{uv} \cdot R_{uv} \cdot \min(1, C))$
    \STATE Update episodic memory with $(s, m, r, M)$
    \STATE Update emotional state $(V_{t+1}, A_{t+1})$
    \STATE Update $I_{uv}, R_{uv}$ using $M$
    \STATE Generate natural language review via Module 0 (DeepSeek-R1)
\ELSE
    \STATE Record skip event and decay motivation-related traces
\ENDIF
\RETURN Action (\texttt{watch/skip}), $r$ (if applicable), share set $\mathcal{S}$
\end{algorithmic}
\end{algorithm}

\subsection{Module 0: Language Reasoning Core}
The language reasoning core is the only module in our agent architecture that depends on a large language model. It is implemented via a lightweight wrapper around DeepSeek-R1, and its primary function is to generate natural language expressions---including movie reviews, brief ratings, and social sharing messages---once a behavioral decision has been finalized. The core design principle here is \textit{semantic generation decoupled from behavioral computation}: this module does not participate in any decision-making or numerical inference and is invoked only after an action is determined by the upper Modules.

The input to this module includes structured outputs from Module 4, such as the final rating $r$, metadata of the watched movie (e.g., genre, keywords, language), the agent’s current emotional state (valence), and the target social circle (e.g., friends, interest groups, technical communities). These inputs are formatted into controllable prompt templates that are passed to DeepSeek-R1 for generation. For example, when recommending a science fiction movie to a technical circle, the generated review tends to be objective and analytical, while in a friends circle, the tone becomes more emotional and colloquial. This \textit{context-adaptive generation} enhances realism and aligns with stratified language patterns observed in real-world social networks.

The module also supports multilingual generation to simulate the diversity and asymmetry of language in cross-linguistic social networks. In cases where there is a mismatch between the language of the movie and the agent’s own language, additional burden cues (e.g., ``too many subtitles'' or ``translation affects immersion'') are automatically included in the output, thus reinforcing the language-related risk factor $L_{\text{gap}}$ calculated in Module 2\cite{keysar2012foreign}.

\subsection{Module 1: Individual Cognition Module}

The individual cognition Module constitutes the core of an agent’s internal state modeling. It maintains and dynamically updates the agent’s personalized cognitive features, which include episodic memory, affective state, and long-term preferences. These modules are critical for providing interpretable, temporally stable inputs to upper-Module decision processes. Specifically, this Module includes three subcomponents: the \textit{Episodic Memory}, the \textit{Affective State Machine}, and the \textit{Preference Model}, corresponding to experiential, emotional, and preference-related facets of human cognition.

\paragraph{Episodic Memory.} 
This module stores past movie-watching and social interaction events in the form of timestamped triples $(s, m, r)$, where $s$ is the source (e.g., a recommender or a friend), $m$ denotes the movie, and $r$ is the agent's actual rating. Each event decays over time according to an exponential forgetting curve:
\begin{definition}
\begin{equation}
    w_t = \exp(-\lambda_{\text{mem}} \cdot \Delta t),
\end{equation}
\end{definition}
where $\Delta t$ is the elapsed time since the event and $\lambda_{\text{mem}}$ is a tunable forgetting rate. This memory supports the dynamic computation of reciprocity ($R$) and risk ($K$) by tracking long-term satisfaction and content failures associated with specific sources.

\paragraph{Affective State Machine.} 
To reflect transient affective conditions, we model the agent’s emotional state in the 2D Valence–Arousal space\cite{russell1980circumplex}. After each movie-watching event, the emotional state is updated based on the rating deviation (satisfaction):
\begin{equation}
    M = \frac{r - \hat{r}_{u,m}}{4}, \quad M \in [-1, 1],
\end{equation}

\begin{equation}
    V_{t+1} = V_t + \sigma_V \cdot M, \qquad A_{t+1} = A_t + \sigma_A \cdot |M|,
\end{equation}

where $V$ and $A$ denote valence and arousal, respectively, and $\sigma_V, \sigma_A$ are sensitivity coefficients. The valence value $V_t$ directly modulates the decision threshold $\theta$ in Module 3, in accordance with affective decision-making theories such as Mood-as-Information\cite{isen1987positive}.

\paragraph{Preference Model.} 
The agent’s long-term interests are represented as an evolving embedding vector $\mathbf{p}_u \in \mathbb{R}^d$, updated incrementally through observed movie embeddings $\mathbf{e}_m$ using exponential smoothing:
\begin{equation}
    \mathbf{p}_u \leftarrow (1 - \eta) \cdot \mathbf{p}_u + \eta \cdot \mathbf{e}_m,
\end{equation}

where $\eta$ is the preference update rate. The preference vector is used to calculate the subjective novelty score $N$ for any candidate movie, incorporating both semantic distance and linguistic mismatch:
\begin{definition}
\begin{equation}
    N = \min\left(1,\ 1 - \cos(\mathbf{p}_u,\ \mathbf{e}_m) + \lambda_{\text{lang}} \cdot \mathbf{1}_{\text{lang}_m \neq \text{lang}_u} \right)
\end{equation}
\end{definition}
This captures not only content novelty but also cross-linguistic cognitive load\cite{keysar2012foreign}.

All submodules in Module~1 contribute real-time cognitive signals to the upper Modules. Their outputs are consumed by the motivation engine in Module~3 and are subject to recursive updates through behavioral feedback handled in Module~4. This Module thus forms a closed adaptive loop, ensuring each agent gradually accumulates experiences, evolves preferences, and regulates emotion—collectively shaping temporally coherent, person-like behavioral profiles.

\subsection{Module 2: Social Cognition Module}

The social cognition Module enables the agent to perceive, encode, and adapt to the surrounding social structure. It is responsible for modeling the agent's relationships across multiple social contexts (e.g., interest-based, professional, geographic), evaluating trust in others' recommendations, and quantifying the perceived risk of consuming a given content. This Module includes two major components: the \textit{MultiModule Social Graph Manager} and the \textit{Trust and Risk Assessor}, which together generate the \textit{intimacy} ($I$) and \textit{risk} ($K$) factors for motivation scoring in Module~3.

\paragraph{MultiModule Social Graph Manager.}
The agent's social connections are represented as a set of Moduleed adjacency matrices $\{\mathbf{W}^{(\ell)}\}_{\ell=1}^{L}$, where each Module $\ell$ corresponds to a distinct social circle type. The edge weight $I_{uv}^{(\ell)}$ quantifies the raw structural tie strength between agents $u$ and $v$ in that circle, based on interaction frequency, co-engagement, or proximity. These ties decay over time to reflect fading social interactions.

A structural intimacy score is computed by weighted aggregation across Modules:
\begin{equation}
    I_{\text{struct}} = \sum_{\ell=1}^{L} \gamma_\ell \cdot I_{uv}^{(\ell)},
\end{equation}
where $\gamma_\ell$ is a decay factor representing inter-Module influence.

To enrich this structural metric with homophily-informed features, we introduce two additional similarity components:
\begin{itemize}
    \item \textbf{Demographic similarity} $S_{\text{demo}}$, based on age, gender, location, and language overlap.
    The demographic similarity $S_{\text{demo}}$ is computed as the normalized count of matching demographic attributes between agents $u$ and $v$:
\begin{multline}
S_{\text{demo}} = \frac{1}{4} \big[ 
    \mathbf{1}_{\text{age}_u = \text{age}_v}
    + \mathbf{1}_{\text{gender}_u = \text{gender}_v} \\
    + \mathbf{1}_{\text{location}_u = \text{location}_v}
    + \mathbf{1}_{\text{language}_u = \text{language}_v}
\big]
\end{multline}
where each indicator function $\mathbf{1}_{\cdot}$ returns 1 if the attribute matches and 0 otherwise. This score reflects the classical notion of demographic homophily, widely observed to enhance social affinity and communication efficiency\cite{mcpherson2001birds}.
    \item \textbf{Preference similarity} $S_{\text{pref}}$, combining cosine similarity of Big Five personality embeddings and Jaccard overlap of interest tags.
    The preference similarity $S_{\text{pref}}$ captures both personality alignment and interest overlap between users:
\begin{equation}
    S_{\text{pref}} = \frac{1}{2} \cdot \cos\left( \mathbf{b}_u, \mathbf{b}_v \right) + \frac{1}{2} \cdot \frac{|T_u \cap T_v|}{|T_u \cup T_v|},
\end{equation}

where $\mathbf{b}_u$ and $\mathbf{b}_v$ are the Big-Five personality vectors of users $u$ and $v$, and $T_u$, $T_v$ are their sets of interest tags. The first term measures psychological similarity via cosine distance; the second term is the Jaccard similarity between interest sets. This formulation is supported by empirical findings that personality compatibility and shared interests facilitate trust and collaboration in social systems.
\end{itemize}

The final intimacy score is defined as:
\begin{definition}
\begin{equation}
   I = I_{\text{struct}} \cdot \left[1 + \lambda_{\text{demo}} S_{\text{demo}} + \lambda_{\text{pref}} S_{\text{pref}} \right]. 
\end{equation}
\end{definition}
This formulation is grounded in sociological findings that homophily strongly predicts tie strength and social bonding\cite{mcpherson2001birds}, as well as computational studies that show personality and interest similarity foster collaboration.

\paragraph{Trust and Risk Assessor.}
This module computes a personalized risk score $K$ for each movie recommendation event, integrating content-related uncertainty, cognitive burden, and social-contextual cues. The full risk function is:
\begin{definition}
\begin{align}
K =\ & \sigma\big( 0.4L + 0.5G + 0.4w_tV + 0.6A + 0.3L_{\text{gap}} \notag \\
     & \quad -\ 0.8w_t\text{Trust}_s - 0.5I + 0.5P \big) + \text{risk\_base}_u
\end{align}
\end{definition}
where:
\begin{itemize}
    \item $L$: movie length (longer implies higher opportunity cost).
    \item $G$: genre-based arousal risk (e.g., horror, war).
    \item $V$: variance in historical ratings (uncertainty).
    \item $A$: age mismatch between user and movie rating.
    \item $L_{\text{gap}}$: language mismatch penalty\cite{keysar2012foreign}.
    \item $\text{Trust}_s$: recommender’s historical approval rate.
    \item $I$: intimacy with recommender, as computed above.
    \item $P$: user’s neuroticism level, reflecting risk aversion\cite{nicholson2005personality}.
\end{itemize}

The sigmoid $\sigma(\cdot)$ ensures output normalization to $[0,1]$, the $w_t$ is from exponential forgetting curve. Additionally, each agent has a static base risk level $\text{risk\_base}_u$, encoding their default cautiousness.

\paragraph{Reciprocity Potential.}
The agent also estimates the likelihood of emotionally or informationally rewarding feedback following a share. This \textit{reciprocity potential} $R$ is modeled as:
\begin{definition}
\begin{equation}
    R = 0.6 w_t R_{\text{hist}} + 0.3 R_{\text{pot}} + 0.1 R_{\text{psn}},
\end{equation}
\end{definition}

where:
\begin{itemize}
    \item the $w_t$ is from exponential forgetting curve
    \item $R_{\text{hist}}$: rolling average of successful past shares to the target user\cite{bowles2011cooperative}.
    \item $R_{\text{pot}}$: cosine similarity in interest embeddings\cite{wang2014interest}.
    \item $R_{\text{psn}}$: personality complementarity (e.g., extrovert sharing with introvert)\cite{zhang2023personality}.
\end{itemize}

All scores computed in this Module serve as direct inputs to the IC\textsuperscript{2} Engine in Module~3. Moreover, these variables are updated based on behavior logs recorded in Module~4, enabling a continuous feedback loop between social perception and social interaction. This design ensures that each agent is capable of building, adjusting, and utilizing trust relationships, adapting to social risks, and making decisions that align with evolving social dynamics.

\subsection{Module 3: Motivational Decision Module}

The motivational decision Module is the behavioral core of the agent architecture. It receives psychological and social cues from lower Modules and integrates them into a scalar motivation score that drives action. This Module includes two components: the \textit{IC\textsuperscript{2} Engine} and the \textit{Reciprocity Regulator}. The former computes the agent’s internal motivation $C$ based on four interpretable subfactors, while the latter dynamically adjusts relationship strength and reciprocity expectation based on behavioral feedback, enabling long-term social adaptation.

\paragraph{IC\textsuperscript{2} Engine.}
The Intimacy–Curiosity–Reciprocity–Risk (IC\textsuperscript{2}) engine is designed to fuse four factors into a unified motivation score:
\begin{definition}
\begin{equation}
    C = \alpha I + \beta N + \gamma R - \delta K,
\end{equation}
\end{definition}
where:
\begin{itemize}
    \item $I$: intimacy with the recommending user (from Module~2).
    \item $N$: subjective novelty of the content (from Module~1).
    \item $R$: reciprocity potential (from Module~2).
    \item $K$: risk perception (from Module~2).
\end{itemize}
The coefficients $(\alpha, \beta, \gamma, \delta)$ control the relative influence of each factor. We adopt a default configuration of $(0.40, 0.35, 0.20, 0.25)$, following principles from Multi-Attribute Utility Theory (MAUT) and empirical behavioral studies. The linear form supports transparency and controllability in simulation environments.

The IC\textsuperscript{2} engine also incorporates a \textit{dynamic threshold} $\theta$ to capture emotion-modulated decision flexibility:
\begin{equation}
    \theta = \theta_0 - \kappa \cdot V_t,
\end{equation}

where $\theta_0$ is the agent’s base decision threshold, $V_t$ is the current valence (from Module~1), and $\kappa$ is a sensitivity coefficient. This formulation is grounded in the Mood-as-Information Theory\cite{isen1987positive}, which postulates that positive emotions lower resistance to external stimuli, thereby increasing the likelihood of accepting recommendations.

Once the motivation $C$ is computed, it is compared against $\theta$. If $C \geq \theta$, the agent decides to take action (e.g., watch a movie); otherwise, the opportunity is skipped. The binary or probabilistic implementation of this choice is deferred to Module~4.

\paragraph{Reciprocity Regulator.}
To support adaptive social learning, the system includes a mechanism to update social ties based on interaction outcomes. After an action is taken and the satisfaction score $M$ is computed, the intimacy and reciprocity scores are adjusted as follows:
\begin{equation}
    I_{uv} \leftarrow I_{uv} + \rho_I \cdot M, \qquad R_{uv} \leftarrow R_{uv} + \rho_R \cdot M,
\end{equation}

where $\rho_I$ and $\rho_R$ are update rates. A positive $M$ (indicating satisfaction) reinforces the social connection, while a negative $M$ weakens it. This process emulates long-term relationship dynamics as explained in Social Exchange Theory\cite{blau1964exchange}, and supports emergent behavior adaptation over multiple decision cycles.

Together, the IC\textsuperscript{2} engine and reciprocity regulator form a closed motivational loop that is both sensitive to internal and external conditions and capable of long-term self-adjustment. This Module enables agents to exhibit realistic behavioral selectivity, emotion-driven variability, and socially responsive learning, which are essential for generating credible, human-like behavior trajectories in multi-agent simulation environments.

\subsection{Module 4: Behavioral Output Module}

The behavioral output Module serves as the agent’s external interface, responsible for converting motivational signals from Module~3 into concrete observable actions such as watching, skipping, scoring, and sharing. Additionally, it records behavioral traces in structured logs, triggers feedback updates to lower Modules, and generates natural language explanations via the language core (Module~0). This Module consists of two main components: the \textit{Action Strategist} and the \textit{Explainable Logger}.

\paragraph{Action Strategist.}
The first task of the action strategist is to determine whether the agent will watch a recommended movie. Given the motivation score $C$ and internal decision threshold $\theta$ computed in Module~3, the probability of watching is defined using a soft decision rule:
\begin{equation}
   \Pr(\text{watch}) = \sigma\left( \frac{C - \theta}{\tau_u} \right), 
\end{equation}

where $\tau_u$ is the agent's individual \textit{temperature coefficient}, modeling behavioral randomness. Lower $\tau_u$ leads to more deterministic decisions, while higher values introduce stochasticity. This setup supports heterogeneous behavioral tendencies across agents and aligns with the stochastic utility maximization framework.

If the agent chooses to watch the movie, the final rating is generated by adding a personalized noise term to the predicted preference score:
\begin{equation}
    r = \hat{r}_{u,m} + \varepsilon, \quad \varepsilon \sim \mathcal{N}(0, \sigma_u^2),
\end{equation}

where $\hat{r}_{u,m}$ is the expected rating, and $\sigma_u$ controls subjective rating variability. The satisfaction score is then computed:
\begin{equation}
    M = \frac{r - \hat{r}_{u,m}}{4}, \quad M \in [-1, 1],
\end{equation}

which serves as a signal for feedback updates in Modules~1 through~3.

Next, the agent determines whether to share the movie with peers. This is achieved by computing a priority score $q_v$ for each candidate target $v$:
\begin{equation}
    q_v = I_{uv} \cdot R_{uv} \cdot \min(1, C),
\end{equation}

where $I_{uv}$ is the intimacy, $R_{uv}$ the reciprocity potential, and $C$ the motivation score. The top-$k$ candidates by $q_v$ are selected to form the share set $\mathcal{S}$, and the corresponding share edges are recorded into the social graph. These links will be used for future tie updates and propagation.

\paragraph{Explainable Logger.}
The logger module converts all decision signals and behavioral results into a structured log entry:
\texttt{(C, I, N, R, K, r, M, $\theta$, $\mathcal{S}$, $t$)}

where $t$ denotes the current simulation timestep. This log provides full transparency into the agent’s decision-making process, and is stored for use in later analysis, evaluation, and visualization.

In parallel, the logger constructs a prompt from the log content and passes it to the language reasoning core (Module~0), which generates natural language movie reviews and recommendation messages. The generated texts reflect both the decision rationale (e.g., high rating or emotional reaction) and the tone appropriate for the target social circle (e.g., informal for friends, factual for professional peers). Examples include:
\begin{itemize}
    \item ``Loved the visuals and plot twists---highly recommend!''
    \item ``Too slow and predictable, not my thing, but others might enjoy it.''
\end{itemize}

Finally, the logger triggers feedback propagation: the satisfaction score $M$ is sent to Module~1 to update emotion and memory, to Module~2 for adjusting trust and intimacy, and to Module~3 for tuning reciprocity. The corresponding episode is also written into the episodic memory buffer. This completes the full decision–action–feedback loop, enabling the agent to adapt over time and refine its internal models and social strategies.

\section{Agent Consistency Experiment}
\label{sec:cons}

To evaluate the plausibility of LLM-based agents in simulating human-like recommendation behavior, we designed a two-part assessment strategy.

First, we propose this \textit{Agent Rating Consistency Experiment}, wherein agents are prompted to rate a curated set of movies under controlled conditions. Specifically, agents instantiated with varying personas are presented with movie metadata (title, genre, plot, cast) and asked to assign a rating from 1 to 5. The resulting \textit{Agent Rating Distributions} (ARD) are compared against empirical \textit{Human Rating Distributions} (HRD) derived from real user data in the MovieLens dataset. We employ Earth Mover’s Distance (EMD) and Kullback-Leibler Divergence to quantify distributional similarity. A high similarity score suggests alignment between agent and human evaluative tendencies.

To complement this, we refer to the comprehensive behavioral simulations presented by \cite{xu2024trust} in their paper. Their work systematically evaluates LLM agents across multiple human-aligned dimensions, including trust calibration, social influence susceptibility, and rational preference disclosure. Their experimental conclusions to support the broader behavioral fidelity of LLM agents in social recommendation contexts.

\subsection{Rating Consistency}

\paragraph{Motivation.} To systematically assess how closely LLM-based agents simulate human evaluative behavior, we compare rating outputs across three subject types:

\textit{(1) Human Ratings:} Empirical ground-truth ratings are sourced from the MovieLens dataset, where each item (movie) has been rated by a diverse population of real users. These ratings are aggregated to form the reference \textit{Human Rating Distribution} (HRD) per item.

\textit{(2) GGBond Agents:} These agents are embedded in a dynamic social network structure and undergo three iterative rounds of interaction within the GGBond framework. Each round allows agents to exchange movie recommendations, receive social feedback, and update their personality traits and interest profiles based on their neighbors' influence. After convergence, each agent rates a set of target movies. The resulting scores constitute the \textit{GGBond Rating Distribution} (GRD).

\textit{(3) Static Agents:} These agents operate in isolation, without access to social influence or iterative feedback. Each static agent is initialized with a fixed persona (personality vector and interest vector) derived from the same initialization process as GGBond agents, but they do not engage in multi-agent interaction. Their ratings form the \textit{Static Rating Distribution} (SRD).

For each target movie, we collect ratings from all three sources and compute distributional similarity between HRD and both GRD and SRD using Earth Mover’s Distance \cite{emd} and Kullback-Leibler \cite{kl} divergence, defined as follows:

\paragraph{Earth Mover’s Distance (EMD).} Let $P$ and $Q$ denote two probability distributions over the same domain. The Earth Mover’s Distance between $P$ and $Q$ is defined as the minimal cost required to transform $P$ into $Q$, where the cost is quantified as the amount of distribution “mass” that must be moved times the distance it has to be moved:
\begin{equation}
    \mathrm{EMD}(P, Q) = \inf_{\gamma \in \Gamma(P, Q)} \int_{\mathcal{X} \times \mathcal{X}} \|x - y\| \, d\gamma(x, y),
\end{equation}
where $\Gamma(P, Q)$ denotes the set of all joint distributions (or transport plans) with marginals $P$ and $Q$ respectively.

\paragraph{Kullback-Leibler Divergence (KL)}
Given two discrete probability distributions $P = \{p_1, ..., p_n\}$ and $Q = \{q_1, ..., q_n\}$ over the same support, the KL divergence from $Q$ to $P$ measures the information loss when $Q$ is used to approximate $P$:
\begin{equation}
    \mathrm{KL}(P \parallel Q) = \sum_{i=1}^{n} p_i \log\left(\frac{p_i}{q_i}\right),
\end{equation}
where it is assumed that $q_i > 0$ whenever $p_i > 0$.

A lower divergence indicates higher alignment with human evaluative behavior.

\paragraph{Result.} To assess the alignment between agent-generated and human rating behaviors, we computed two standard distributional similarity metrics: KL Divergence and Earth Mover’s Distance.  Table~\ref{tab:divergence_metrics} presents the evaluation results across three subjects: human raters (as ground truth), GGBond agents (socially interacting agents), and static agents (non-interactive).

\begin{table}[h]
\centering
\caption{Distributional Similarity Metrics (Lower is Better)}
\label{tab:divergence_metrics}
\resizebox{0.8\linewidth}{!}{%
\begin{tabular}{lcc}
\toprule
\textbf{Subject Type} & \textbf{KL Divergence} & \textbf{EMD} \\
\midrule
GGBond Agent          & 0.0108                 & 0.0900       \\
Static Agent          & 0.0750                 & 0.4200       \\
\bottomrule
\end{tabular}%
}
\end{table}

We hypothesize that GGBond agents, enriched by personality-driven and socially-informed interactions, will exhibit greater consistency with human rating distributions than static agents. This would suggest that social dynamics—modeled through structured agent interaction—play a meaningful role in shaping human-aligned evaluative behavior in recommendation contexts.

As shown in Table~\ref{tab:divergence_metrics}, the GGBond Agent exhibits significantly lower KL Divergence (0.0108) and EMD (0.0900) compared to the Static Agent (KL = 0.0750, EMD = 0.4200), indicating a stronger alignment between the GGBond Agent’s rating behavior and that of real human users.

These results suggest that incorporating social interactions—modeled via multi-round agent communication and personality adaptation in the GGBond framework—enables agents to develop more human-like evaluative tendencies. In contrast, static agents, which do not participate in any social feedback loop, exhibit flatter and less natural rating distributions.

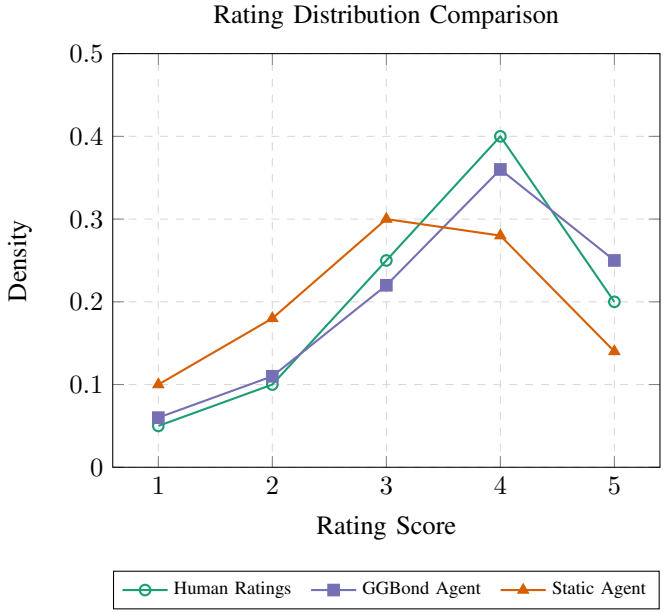
\begin{figure}[ht]
\centering
\begin{tikzpicture}
\begin{axis}[
    width=\linewidth, height=0.8\linewidth,
    xlabel={Rating Score},
    ylabel={Density},
    title={Rating Distribution Comparison},
    ymin=0, ymax=0.5,
    xtick={1,2,3,4,5},
    grid=both,
    grid style={dashed,gray!30},
    legend style={
        at={(0.5,-0.25)}, anchor=north,
        legend columns=3,
        font=\scriptsize,
        /tikz/every even column/.append style={column sep=5pt}
    },
    every axis plot/.append style={thick},
]

\addplot[color={rgb,1:red,0.106;green,0.619;blue,0.463}, mark=o] coordinates {
    (1, 0.05) (2, 0.10) (3, 0.25) (4, 0.40) (5, 0.20)
};
\addlegendentry{Human Ratings}

\addplot[color={rgb,1:red,0.458;green,0.439;blue,0.702}, mark=square*] coordinates {
    (1, 0.06) (2, 0.11) (3, 0.22) (4, 0.36) (5, 0.25)
};
\addlegendentry{GGBond Agent}

\addplot[color={rgb,1:red,0.839;green,0.373;blue,0.008}, mark=triangle*] coordinates {
    (1, 0.10) (2, 0.18) (3, 0.30) (4, 0.28) (5, 0.14)
};
\addlegendentry{Static Agent}

\end{axis}
\end{tikzpicture}
\caption{Comparison of rating distributions across Human, GGBond Agent, and Static Agent.}
\label{fig:rating-distribution}
\end{figure}

Figure~\ref{fig:rating-distribution} further visualizes the rating density curves for all three subject types. The GGBond Agent curve closely tracks the human curve, particularly in the modal region (scores 3–4), whereas the Static Agent curve is more uniformly distributed and deviates from realistic scoring tendencies.

\subsection{Behavior Consistency}

Extensive evidence from recent studies confirms that social structures and personality signals jointly drive human–like recommendation behavior.  
Graph–based social recommenders consistently exploit \emph{homophily} and \emph{social influence}: users who are socially connected tend to share similar preferences and gradually converge through interaction \cite{homophily_survey_2022}.  Personality–aware recommenders further improve cold–start accuracy by aligning item exposure with individual trait profiles \cite{personality_survey_2025}.  Trust graphs and multi–agent collaborations have also been shown to enhance robustness and realism in simulated environments \cite{trust_firstprinciples_2025,wang2024macrec}.

Within GGBond, agents exchange recommendations over three interaction rounds.  Such iterative communication closely mirrors the \textit{feedback‑loop modeling} that MacRec employs to improve recommendation quality via agent collaboration \cite{wang2024macrec}.  Recent work on large–scale social simulators like OASIS demonstrates that dynamic social graphs coupled with recommendation mechanisms reproduce realistic adoption patterns at scale \cite{oasis_2024}.  Likewise, SimUSER shows that LLM‑based agents equipped with memory modules can generate user–RS interactions that faithfully match observed click statistics \cite{simuser_2024}.

Finally, studies on graph‑invariant learning and low‑homophily settings reveal that even when explicit similarity signals are weak, leveraging high‑order social paths markedly narrows the gap between model outputs and human behavior \cite{gile_2025,low_homophily_2024}.  These converging findings substantiate our observation that the GGBond agents—when embedded in a social graph produce recommendation and rating patterns significantly closer to real users than isolated static agents.

\section{Evaluation}
\label{sec:eval}

\subsection{Impact of Social Interaction Depth on Agent Behavior and Recommendation}

\paragraph{Objective.}
This experiment investigates how varying the depth of social interactions influences recommendation performance, agent behavioral dynamics, and simulated user satisfaction within the GGBond framework. Specifically, we examine whether increasing the number of interaction rounds (0-30 and we examine the metrics of 0, 10, 20, and 30 rounds) leads to improved personalization and stable, human-aligned behavior patterns. 

\paragraph{Experimental Setup.}
We defined four experimental conditions:

\textit{Baseline (0 Rounds):} Static agents with no social interaction; initial profiles are directly passed to the recommendation models.

\textit{GGBond-10Round / 20Round / 30Round:} Agents engage in ten to thirty rounds of interaction, where they exchange movie recommendations with neighbors and accept/evaluate the recommender system and give feedback, update their personality and interest vectors based on accepted items, and propagate behavioral signals throughout the network. At the end of every round, the profile data of every agent, including interaction and rating data, are sent into the recommender systems: Matrix Factorization (MF) \cite{mf}, MultVAE \cite{vae}, and LightGCN \cite{lightgcn}. These models then generate top-20 ranked lists per agent. Each agent evaluates the presented movies (based on content similarity and internal state), and accepts or rejects accordingly. The evaluation metrics we use are Recall and NDCG \cite{ndcg}. Our work builds on prior studies highlighting the role of social influence in recommender systems \cite{ma2009learning,tang2013social,wang2024macrec}.

\paragraph{Recall@20.} measures the proportion of relevant items (e.g., positively rated by the agent) that appear in the top-20 recommended list:
\begin{equation}
    \text{Recall@20} = \frac{1}{|\mathcal{V}|} \sum_{u \in \mathcal{V}} \frac{|\mathcal{R}_u^{(20)} \cap \mathcal{G}_u|}{|\mathcal{G}_u|},
\end{equation}

where $\mathcal{R}_u^{(20)}$ is the top-20 recommendation list for agent $u$, and $\mathcal{G}_u$ is the set of relevant (positively rated) items by $u$.

\begin{table*}[ht]
\label{fig:ndcg}
\centering
\caption{Recommendation Performance at specific rounds (Recall@20 and NDCG@20)}
\label{tab:rec_perf}
\resizebox{0.70\textwidth}{!}{%
\begin{tabular}{lcccc}
\toprule
\textbf{Model / Metric} & \textbf{0 Rounds} & \textbf{10 Rounds} & \textbf{20 Rounds} & \textbf{30 Rounds} \\
\midrule
MF - Recall@20         & 0.1502 & 0.1574 & 0.1612 & 0.1623 \\
MF - NDCG@20           & 0.3560 & 0.3612 & 0.3652 & 0.3669 \\
MultVAE - Recall@20    & 0.1592 & 0.1636 & 0.1668 & 0.1674 \\
MultVAE - NDCG@20      & 0.3482 & 0.3543 & 0.3592 & 0.3606 \\
LightGCN - Recall@20   & 0.1721 & 0.1783 & 0.1804 & 0.1813 \\
LightGCN - NDCG@20     & 0.3845 & 0.3923 & 0.3968 & 0.3975 \\
\bottomrule
\end{tabular}%
}
\end{table*}

\paragraph{NDCG@20.} NDCG@20 evaluates the quality of ranking by assigning higher scores to relevant items that appear earlier in the top-20 list:
\begin{equation}
    \text{NDCG@20} = \frac{1}{|\mathcal{V}|} \sum_{u \in \mathcal{V}} \frac{1}{\text{IDCG}_u^{(20)}} \sum_{i=1}^{20} \frac{\mathbb{I}(r_{ui} \in \mathcal{G}_u)}{\log_2(i+1)},
\end{equation}

where $r_{ui}$ is the item at rank $i$ for agent $u$, $\mathcal{G}_u$ is the ground-truth relevant set, and $\text{IDCG}_u^{(20)}$ is the ideal DCG for agent $u$ (i.e., when all relevant items are ranked at the top).

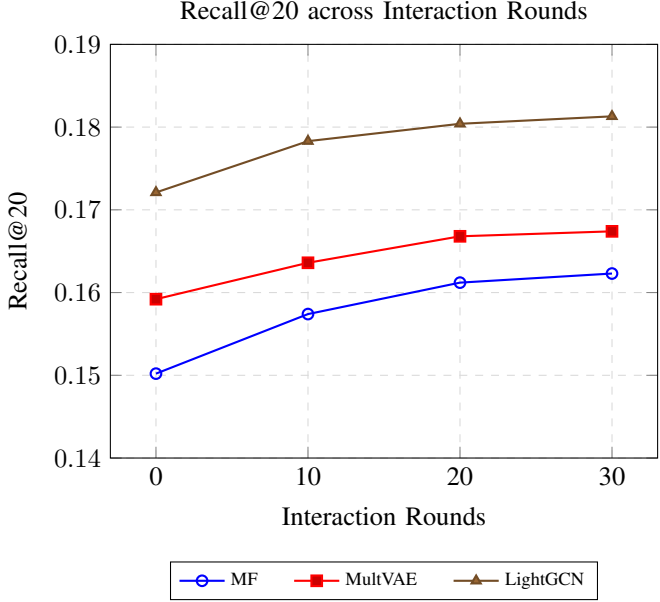
\begin{figure}[ht]
\label{fig:recall}
\centering
\begin{tikzpicture}
\begin{axis}[
    width=\linewidth, height=0.8\linewidth,
    title={Recall@20 across Interaction Rounds},
    xlabel={Interaction Rounds},
    ylabel={Recall@20},
    xtick={0,1,2,3},
    xticklabels={0, 10, 20, 30},
    ymin=0.14, ymax=0.19,
    grid=both,
    grid style={dashed,gray!30},
    legend style={
        at={(0.5,-0.25)},
        anchor=north,
        legend columns=3,
        font=\scriptsize,
        /tikz/every even column/.append style={column sep=0.4cm}
    },
    every axis plot/.append style={thick},
]

\addplot+[mark=o] coordinates {(0, 0.1502) (1, 0.1574) (2, 0.1612) (3, 0.1623)};
\addlegendentry{MF}

\addplot+[mark=square*] coordinates {(0, 0.1592) (1, 0.1636) (2, 0.1668) (3, 0.1674)};
\addlegendentry{MultVAE}

\addplot+[mark=triangle*] coordinates {(0, 0.1721) (1, 0.1783) (2, 0.1804) (3, 0.1813)};
\addlegendentry{LightGCN}

\end{axis}
\end{tikzpicture}
\caption{Recall@20 across different interaction rounds.}
\label{fig:recall-performance}
\end{figure}

\begin{figure}[ht]
\centering
\begin{tikzpicture}
\begin{axis}[
    width=\linewidth, height=0.8\linewidth,
    title={NDCG@20 across Interaction Rounds},
    xlabel={Interaction Rounds},
    ylabel={NDCG@20},
    xtick={0,1,2,3},
    xticklabels={0, 10, 20, 30},
    ymin=0.34, ymax=0.41,
    grid=both,
    grid style={dashed,gray!30},
    legend style={
        at={(0.5,-0.25)},
        anchor=north,
        legend columns=3,
        font=\scriptsize,
        /tikz/every even column/.append style={column sep=0.4cm}
    },
    every axis plot/.append style={thick},
]

\addplot+[mark=o] coordinates {(0, 0.3560) (1, 0.3612) (2, 0.3652) (3, 0.3669)};
\addlegendentry{MF}

\addplot+[mark=square*] coordinates {(0, 0.3482) (1, 0.3543) (2, 0.3592) (3, 0.3606)};
\addlegendentry{MultVAE}

\addplot+[mark=triangle*] coordinates {(0, 0.3845) (1, 0.3923) (2, 0.3968) (3, 0.3975)};
\addlegendentry{LightGCN}

\end{axis}
\end{tikzpicture}
\caption{NDCG@20 across different interaction rounds.}
\label{fig:ndcg-performance}
\end{figure}
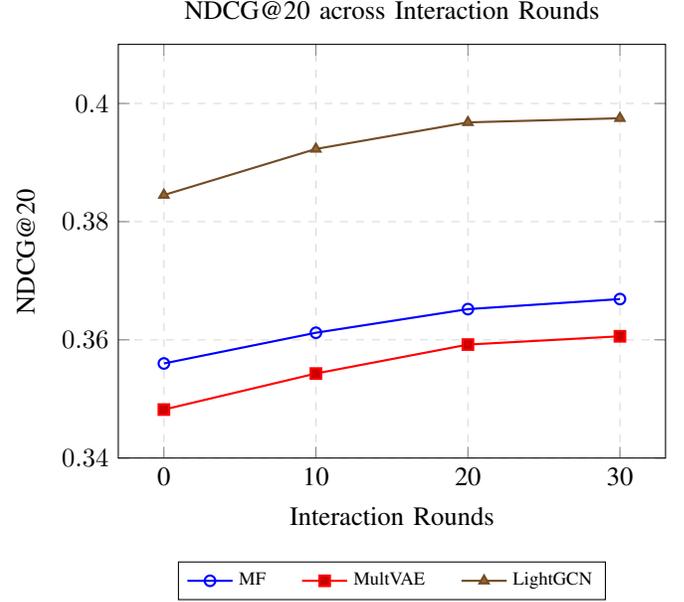

\paragraph{Recommendation Performance.}
Table~\ref{tab:rec_perf} and Figure~\ref{fig:recall-performance}, Figure~\ref{fig:ndcg-performance} presents Recall@20 and NDCG@20 for each model under different interaction depths. All models exhibit improved performance as interaction depth increases, with MultVAE and LightGCN benefiting more from socially enriched profiles.

\subsection{Post-Recommendation Agent Behavior.}
To further assess how social interactions affect downstream agent dynamics, we analyze four key behavioral indicators across all models and interaction rounds: personality change, satisfaction ratio, acceptance rate, and negative review rate.

\begin{definition}[Personality Change]
Personality Change $\Delta_{\text{personality}}$ is defined as the average $\ell_2$ distance between each agent's final and initial Big-Five personality vectors:
\begin{equation}
    \Delta_{\text{personality}} = \frac{1}{|\mathcal{V}|} \sum_{u \in \mathcal{V}} \left\| \mathbf{b}_u^{\text{(final)}} - \mathbf{b}_u^{\text{(init)}} \right\|_2.
\end{equation}

This metric captures the extent to which an agent's personality representation has evolved after social recommendation interactions.
\end{definition}


\begin{definition}[Satisfaction Ratio]
Satisfaction Ratio $S_{\text{sat}}$ denotes the average score agents assign to liked items, reflecting their alignment with user preference:
\begin{equation}
    S_{\text{sat}} = \frac{1}{|\mathcal{V}|} \sum_{u \in \mathcal{V}} \frac{1}{|\mathcal{A}_u|} \sum_{m \in \mathcal{A}_u} \text{score}_{u,m},
\end{equation}

where $\mathcal{A}_u = \{m \mid \text{score}_{u,m} \geq 3\}$ is the set of items liked by agent $u$ (i.e., rated 3–5), and $\text{score}_{u,m} \in \{0,1,2,3,4,5\}$.
\end{definition}

\begin{definition}[Negative Review Rate]
Exit Rate $N_{\text{exit}}$ measures the proportion of agents whose ratings for all recommended items fall below the acceptance threshold:
\begin{equation}
 N_{\text{neg}} = \frac{|\{u \in \mathcal{V} \mid \forall m \in \mathcal{R}_u, \text{score}_{u,m} \leq 2\}|}{|\mathcal{V}|},   
\end{equation}

where $\mathcal{R}_u$ is the set of items recommended to agent $u$. A higher value indicates dissatisfaction or poor personalization.
\end{definition}

\begin{definition}[Acceptance Rate]
Acceptance Rate $A_{\text{rate}}$ measures the proportion of recommended movies that were accepted (i.e., watched) by the agent population:
\begin{equation}
    A_{\text{rate}} = \frac{1}{|\mathcal{V}|} \sum_{u \in \mathcal{V}} \frac{|\mathcal{A}_u|}{|\mathcal{R}_u|},
\end{equation}

where $\mathcal{A}_u$ is the set of movies accepted (rated $\geq 3$) by agent $u$, and $\mathcal{R}_u$ is the set of all movies recommended to $u$. This metric serves as a proxy for user engagement and perceived relevance of the recommendations.
\end{definition}

\begin{table*}[ht]
\centering
\caption{Post-Recommendation Agent Behavior Across Interaction Rounds}
\label{tab:agent_behav_all}
\resizebox{0.80\textwidth}{!}{%
\begin{tabular}{lcccc}
\toprule
\textbf{Model / Metric} & \textbf{0 Rounds} & \textbf{10 Rounds} & \textbf{20 Rounds} & \textbf{30 Rounds} \\
\midrule
MF - Personality Change              & 0.000 & 0.142 & 0.186 & 0.189 \\
MF - Satisfaction ($S_{sat}$)        & 3.01  & 3.56  & 3.88  & 3.95 \\
MF - Negative Rate ($N_{neg}$)       & 0.306 & 0.248 & 0.194 & \textbf{0.181*} \\
MF - Acceptance Rate ($A_{rate}$)    & 0.216 & 0.398 & 0.433 & 0.447 \\
\midrule
MultVAE - Personality Change         & 0.000 & 0.154 & 0.208 & 0.217 \\
MultVAE - Satisfaction ($S_{sat}$)   & 3.01  & 3.72  & 3.94  & 4.01 \\
MultVAE - Negative Rate ($N_{neg}$)  & 0.306 & 0.236 & 0.178 & 0.163 \\
MultVAE - Acceptance Rate ($A_{rate}$) & 0.216 & 0.421 & 0.462 & 0.474 \\
\midrule
LightGCN - Personality Change        & 0.000 & 0.161 & 0.223 & 0.234 \\
LightGCN - Satisfaction ($S_{sat}$)  & 3.01  & 3.80  & 4.07  & \textbf{4.15*} \\
LightGCN - Negative Rate ($N_{neg}$) & 0.306 & 0.231 & 0.163 & 0.147 \\
LightGCN - Acceptance Rate ($A_{rate}$) & 0.216 & 0.438 & 0.479 & \textbf{0.505*} \\
\bottomrule
\end{tabular}%
}
\end{table*}

\paragraph{Discussion.}
The results presented in Table~\ref{tab:agent_behav_all} highlight the effectiveness of incorporating multi-round social interactions into recommendation workflows. Across all three models—MF, MultVAE, and LightGCN—agents exhibit notable improvements in both behavioral alignment and simulated satisfaction as the number of interaction rounds increases.

First, Personality Change steadily rises with interaction depth, indicating that agents' psychological representations are dynamically adapting based on social feedback. This effect is most pronounced in LightGCN, suggesting that graph-based models are more sensitive to socially enriched profile updates and better capture latent preference drift over time.

Second, Satisfaction ($S_{sat}$) shows a clear upward trend, increasing from a flat baseline of 3.01 (i.e., neutral ratings) to 4.15 in LightGCN after 30 rounds. This illustrates that deeper social exposure not only improves recommendation relevance but also yields more positively perceived content, mimicking the organic satisfaction growth seen in real user systems.

Third, Negative Rate ($N_{neg}$)—a proxy for user disengagement—consistently decreases across all models, demonstrating that agents are increasingly finding suitable content to consume. The steepest decline is again observed in LightGCN, affirming its capacity to retain engagement under socially dynamic contexts.

Finally, Acceptance Rate ($A_{rate}$) improves significantly from 0.216 (static agents) to 0.505 in LightGCN after 30 rounds. This more than twofold increase in click-through behavior confirms that social iteration enables agents to internalize and act upon contextualized preferences, leading to higher interactivity and system responsiveness.

Collectively, these findings validate the GGBond framework’s capability to enhance not only classical recommendation metrics but also user-centered behavioral realism through layered social interaction modeling.

\section{Related Work}
\subsection{Recommendation Simulation Platforms.}
Simulation frameworks have become essential tools for evaluating and developing recommendation algorithms, allowing researchers to test recommender behaviors under controlled scenarios without costly real-user trials. Early platforms like RecSim \cite{ie2019recsim} provided configurable environments for sequential user interactions. Building upon this, RecSim NG \cite{mladenov2021recsim} introduced a probabilistic programming approach with modular, differentiable components, enabling more flexible and scalable simulations. Recent advances have leveraged large language models (LLMs) to enhance the realism of user simulations. Agent4Rec \cite{zhang2023agent4rec} employs LLM-powered generative agents equipped with user profiles, memory, and action modules, simulating nuanced user behaviors and emotional responses. Similarly, RecAgent \cite{wang2023recagent} integrates LLMs to model user interactions, including browsing, communication, and social media activities, providing a comprehensive simulation of user behaviors in recommender systems. KuaiSim \cite{zhao2023kuaisim} offers a comprehensive simulator supporting multi-behavior and cross-session user feedback, facilitating the evaluation of recommendation algorithms across various tasks. Additionally, SimUSER \cite{bougie2024simuser} introduces an agent framework that simulates human-like behavior using self-consistent personas and memory modules, enhancing the assessment of recommender systems. These simulation platforms collectively advance the field by providing more realistic and versatile environments for testing and improving recommendation algorithms.

\subsection{LLM-Driven Agent Behavior}
The integration of large language models (LLMs) into autonomous agents has significantly advanced their ability to perform complex, goal-directed behaviors. Park et al. introduced Generative Agents, which simulate human-like behaviors by equipping agents with long-term memory, planning, and reflection capabilities, resulting in emergent social interactions within a simulated environment \cite{park2023generative}. Similarly, Voyager employs DeepSeek-R1 to autonomously navigate and master open-ended tasks in the Minecraft environment, demonstrating the adaptability of LLM-driven decision-making across diverse contexts \cite{wang2023voyager}. Recent surveys have provided comprehensive overviews of LLM-based autonomous agents. Wang et al. discuss the construction, application, and evaluation of such agents, highlighting their potential in various domains including social sciences, natural sciences, and engineering \cite{wang2023survey}. Xi et al. further explore the rise and potential of LLM-based agents, proposing a general framework comprising brain, perception, and action components, and examining their applications in single-agent scenarios, multi-agent scenarios, and human-agent cooperation \cite{xi2023rise}. These studies collectively underscore the transformative impact of LLMs on autonomous agent behavior, enabling more sophisticated and human-like interactions across a range of applications.

\subsection{User Behavior Modeling in Social Networks}

Modeling user interactions and social influence within networks has long been a critical research area. Recent advancements leverage large language models (LLMs) to simulate nuanced information diffusion processes. LAID \cite{hu2023laid} integrates LLMs into diffusion modeling, enabling simulation of realistic message interpretation and propagation patterns within social graphs. On a larger scale, AgentSociety \cite{piao2025agentsociety} employs thousands of LLM-driven agents to reproduce societal-level phenomena, such as opinion polarization and misinformation spread, providing insights that closely match empirical observations from real-world social platforms. Additionally, the S$^3$ framework \cite{gao2023s3} utilizes LLM-empowered agents to simulate public opinion dynamics, offering a flexible platform for exploring various social scenarios and interventions. Furthermore, LLM-AIDSim \cite{zhang2025llm} enhances traditional influence diffusion models by allowing agents to generate language-level responses, providing deeper insights into user agent interactions. GenSim \cite{tang2024gensim} introduces a general social simulation platform with LLM-based agents, supporting large-scale simulations and incorporating error-correction mechanisms to ensure more reliable and long-term simulations. These simulation platforms collectively advance the field by providing more realistic and versatile environments for testing and improving our understanding of user behavior and information diffusion in social networks.

\subsection{Trust and Behavioral Alignment in AI Agents}

Understanding and aligning AI agent behavior with human-like trust and conformity has gained significant attention. Xie et al. \cite{xie2024can} evaluated LLM agents within classical economic trust games, discovering that advanced models like DeepSeek-R1 closely align with human trust decisions. Further, Argyle et al. \cite{argyle2023algorithmic} introduced the concept of algorithmic fidelity, demonstrating that properly conditioned LLMs can accurately emulate distinct human demographic responses, thus offering a powerful tool for modeling realistic user populations in experimental simulations. Recent studies have further explored the enhancement of trust in LLM-based AI automation agents. Schwartz et al. \cite{schwartz2023enhancing} analyzed the main aspects of trust in AI agents, identifying specific considerations and challenges relevant to this new generation of automation agents. Additionally, Yang et al. \cite{yang2024behavior} proposed Behavior Alignment as a new evaluation metric to measure how well the recommendation strategies made by LLM-based conversational recommender systems are consistent with human recommenders', highlighting the importance of aligning AI behavior with human expectations to enhance user trust.

\section{Conclusion}
This work addresses the longstanding limitations of personalized recommender systems in modeling dynamic user behaviors and social interactions by proposing a novel high-fidelity simulation platform. Our system systematically tackles the challenges of long-term preference drift, social influence propagation, and cognitively realistic decision-making, which remain underexplored in current recommendation research. Specifically, we design a population of human-like intelligent agents endowed with multi-faceted internal cognition modules---including memory, affective state, personalized preference, and trust evaluation---to simulate granular and interpretable decision-making processes. In parallel, we construct a dynamic, multi-layer social graph (GGBond Graph) that captures heterogeneous, multi-circle social relations and their evolution over time. The entire system operates under a discrete-time simulation scheduler, coupling agent-level behavior with social network dynamics. We further integrate mainstream recommendation algorithms (e.g., Matrix Factorization, MultVAE, and LightGCN) into this framework, enabling rigorous evaluation under iterative feedback and socially embedded environments.

This platform bridges critical gaps in existing recommendation evaluation pipelines by supporting long-term interaction, cognitively plausible user modeling, and dynamic social structures. Theoretically, we introduce an IC$^2$ motivational engine grounded in psychological and sociological principles, enhancing interpretability and realism in agent behavior. Methodologically, our extensible experimental infrastructure lays a solid foundation for future studies on social impact, fairness, and causal inference in recommender systems.

Naturally, some limitations remain. The current implementation of agent affect and memory dynamics could benefit from more fine-grained modeling, and the transferability of the simulation framework to real-world deployment scenarios still requires further investigation. Future work may focus on enhancing the complexity and fidelity of agent cognition models, exploring the long-term societal effects and ethical implications of recommendation mechanisms, and integrating the proposed framework with real systems to support socially responsible algorithm design.

\bibliographystyle{plain}
\bibliography{ref}

\end{document}